\documentclass[review]{elsarticle}

\usepackage{lineno,hyperref,amssymb}
\modulolinenumbers[5]    

\journal{Journal of \LaTeX\ Templates}

\usepackage{amssymb}
\usepackage{amsmath}
\usepackage{latexsym}
\usepackage{epsfig}
\usepackage{graphicx}
\usepackage{color}
\usepackage{fouridx}

\makeatletter

\renewcommand*{\p@subsection}{}

\renewcommand*{\p@subsubsection}{}
\makeatother

\newcommand{\half}{\frac{1}{2}}

\newcommand{\clP}{{\cal P}}
\newcommand{\clL}{{\cal L}}

\numberwithin{equation}{section}

\begin{document}

\begin{frontmatter}

\title{{\small Chaos, Solitons \& Fractals \textbf{142} (2021) 110488 \\
https://doi.org/10.1016/j.chaos.2020.110488}
--------------------------------------------------------------------- \\
\vline \\ \vline \\ 
Anomalous diffusion in umbrella comb}

\author{A. Iomin}
\ead{iomin@physics.technion.ac.il}
\address{Department of Physics, Technion, Haifa, 32000, Israel}

\begin{abstract}
Anomalous transport in a circular comb is studied.
The circular motion takes place for a fixed radius, while radii are
continuously distributed along the circle.
Two scenarios of the anomalous transport are considered.
The first scenario
corresponds to the conformal mapping of a 2D
comb Fokker-Planck equation on the circular comb.
This topologically constraint motion is named umbrella comb model.
In this case, the reflecting boundary conditions are
imposed on the circular motion, while the radial
motion corresponds to geometric Brownian motion
with vanishing to zero boundary conditions on infinity.
The radial diffusion is described by
the log-normal distribution, which corresponds to exponentially fast
motion with the mean squared displacement (MSD) of the order of $e^t$.
The second scenario corresponds to the circular diffusion
with periodic boundary conditions and the outward radial
diffusion with vanishing to zero boundary conditions at infinity.
In this case the radial motion corresponds to normal diffusion.
The circular motion in both scenarios is a superposition
of cosine functions that results in the stationary Bernoulli polynomials
for the probability distributions.

\end{abstract}


\end{frontmatter}

\section{Introduction}\label{sec:int}

\begin{figure}[htbp]
\includegraphics[width=1.0\hsize]{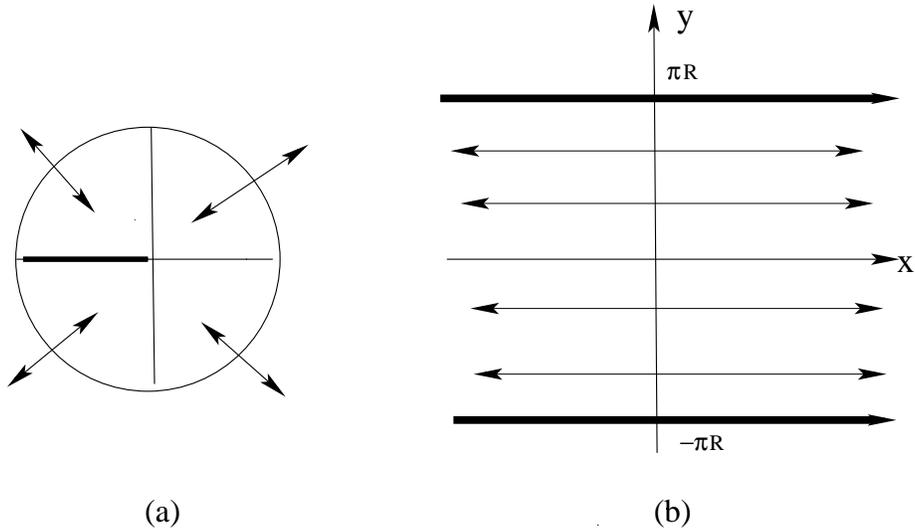}
\caption{Schematic picture of an umbrella comb (a) and its conformal
mapping into the $x-y$ strip comb (b).
The radii are continuously distributed over the circle of the radius
$R$, and the angular motion is possible only at $r=R$. Correspondingly,
the $x$ fingers are continuously distributed along the $y$ backbone.}
\label{fig:fig1}
\end{figure}

In this paper we consider circular and radial motions in combs of
circular geometry, see Fig.~\ref{fig:fig1}, where the radii are
continuously distributed over the circle, and
the circular motion takes place for the
fixed radius $r=R$, only. Fractional diffusion in this geometry
has been studied recently, where both outward and inward radial
diffusion has been considered analytically \cite{FaLiZhLi2019}
and numerically \cite{LiFaLi2020}. Finite time evolution of both
angular and radial probability distribution functions as well as the mean squared displacement have been observed analytically \cite{FaLiZhLi2019} and numerically \cite{FaLiZhLi2019,LiFaLi2020}  for different realizations of the boundary conditions for both angular and radial motions. Further analytical study of the system is important to understand asymptotic transport in the system. We consider two possibilities of boundary conditions for angular diffusions. The first one corresponds to the reflecting boundary condition, and the second one corresponds to the periodic boundary condition.
Our study of anomalous diffusion in this
comb geometry is also motivated by consideration of an
idealized radial transport, which can be also related to
the radial transport model for the Tore Supra tokamak, considered in
Ref. \cite{MiRa2018}.
We however disregard the avalanche dynamics, described by L\'evy flights,
and concentrate our attention to the geometry impact on the
topologically restricted transport in the framework
of the circular comb model, which we call here ``umbrella comb model''.
It is also related to circular anomalous
diffusion in presence of inhomogeneous magnetic fields \cite{Gr2008}.

Anomalous transport in this umbrella comb is described by
a probability distribution function (PDF) $P(r,\phi,t)$
in polar coordinates
to find a particle at the position $(r,\phi)$ at time $t$
in the framework of a Fokker-Planck equation as follows
\begin{subequations}\label{int-uc-1}
\begin{align}
\partial_tP=\Delta P\, , \label{int-uc-1a} \\
\Delta=D_{\phi}\frac{1}{r^2}\partial_{\phi}^2
+D_r\frac{1}{r}\partial_r r\partial_r\, . \label{int-uc-1b}
\end{align}
\end{subequations}
Here $D_{\phi}(r)=D_1\delta(r-R)$ and $D_r=D_r(r)$ are
diffusion coefficients of the
angular and radial directions, respectively. Note that
for the singular $D_{\phi}$, the transport
in the angular direction takes place at $r=R$ only.
The radial diffusion coefficient $D_r(r)$ is a function of the radius,
and this dependence is specified for every scenario
separately.

We consider two scenarios of different realizations of radial diffusion.
In the first scenario we consider geometric Brownian diffusion
along the radii that results from the conformal map of
normal diffusion in $x-y$ plane comb to the circular motion,
as shown in Fig.~\ref{fig:fig1}.
In general case of the conformal map realization, one imposes the
periodic boundary conditions for the circular motion
at $\phi=\pm\pi$, namely $P(R,\phi=\pi,t)=P(R,\phi=-\pi,t)$ for the PDF,
and the shifting boundary conditions for the probability current $\partial_{\phi}P(R,\phi=\pi,t)=-\partial_{\phi}P(R,\phi=-\pi,t)$.
In this case, the amplitudes of the diffusive currents at $\phi=\pm\pi$, clockwise and counterclockwise, are equal to each other. Due to this symmetry at $\phi=\pm\pi$, a cut along the $\phi=\pi$ ray can be performed.
Note that this scenario results from the possible symmetry with respect to the $x$ axis for angular diffusion. It is supported by symmetrical diffusion obtained numerically in Ref. \cite{LiFaLi2020}.
The radial diffusion coefficient is $D_r(r)=D_2r^2$ as the result of the conformal map. We however consider reflecting boundaries at $\phi=\pm\pi$ motivated by dynamical chaos \cite{LiLi83}.
That is, there is an infinite wall\footnote{Note that
this specific choice of the boundary condition can be
replaced by a delta potential, which affects the circular
diffusion like in the Azbel'-Kaner effect \cite{Io94}, or
in chaotic motion of persistent current \cite{Io95}.},
at $\phi=\pm \pi$, or a cut along the
$\phi=\pi$ ray, where $\partial_{\phi}P(R,\phi=\pm \pi,t)=0$.
The second scenario corresponds to the realization of Brownian normal diffusion along the radii, when the radial diffusion coefficient
is taken to be a constant value, $D_r=\mathrm{const}$.
In this case, we consider the periodic boundary condition
at $P(R,\phi=\pi,t)=P(R,\phi=-\pi,t)$.
The zero boundary conditions for the radial directions will be
specified separately for each angular scenario.

In sequel, the section titles are according to the different boundary conditions. However, one should bear
in mind that the difference among these scenarios is due to their
radial diffusivity. That is, the first one is a consequence of the
conformal mapping of the $x-y$ comb, which leads to inhomogeneous
(space dependent) radial diffusivity, while the second one is according to
$r-\phi$ comb constraint with 
a constant diffusion coefficient in the radial direction.

\section{Reflecting boundary conditions}\label{sec:rbc}

In this section, we concentrate our attention on the geometry
impact in the framework of a standard comb model,
which can be mapped onto the circle
and vice versa by the conformal way.
Then the complex plane $w=(r,\phi)$ is mapped on
the complex plane $z=(x,y)$.
The reflecting boundary conditions play important role at this
conformal mapping. Indeed,
if we take a cut along $\phi=\pi$ that yields the reflecting boundaries at
$\phi=\pm\pi$. Then we have $x=\ln (r/R)$ and $y=R\phi$  \cite{LavrShab87};
the map is shown in Fig. \ref{fig:fig1}. The comb model,
which describes anomalous diffusion in the  $x-y$ strip,
shown in Fig.~\ref{fig:fig1}, reads
\begin{equation}\label{rbc-1}
\partial_tP(x,y,t)=D_1\delta(x)\partial_y^2P(x,y,t)
+D_2\partial_x^2P(x,y,t)\, .
\end{equation}
The boundary and the initial conditions are
$P(x=\pm\infty,y,t)=\partial_xP(x=\pm\infty,y,t)=0$,
$\partial_yP(x,y=\pm \pi R,t)=0$, and
$P(x,y,t=0)=\delta(x)\delta(y)$, respectively, and these
conditions reflect the boundaries and the initial condition in the
polar coordinates, as well. The radial motion in the umbrella
comb in Eq. \eqref{int-uc-1a} corresponds to
a dilation(contraction) operator $D_2(r\partial_r)^2$,
which results from the conformal map
$D_2\partial_x^2\rightarrow D_2(r\partial_r)^2$ with the diffusion
coefficient $D_2\rightarrow D_r=D_2r^2$.
This inhomogeneous diffusion
results from the \textit{conformal map}
and corresponds to a multiplicative
white noise and is known as the so-called geometric,
or exponential Brownian motion\footnote{For $r>0$, it describes
\textit{e.g.}, a stock price behavior as a Wiener process for
$x=\log(r)$, which is known as the Black-Scholes model \cite{BlSch73}.
In the present consideration, it can be considered as an exponential
instability of plasma in the radial direction in tokamaks
\cite{MiRa2018}. Note that the dilation operator in dynamical systems relates to an inverted quartic potential, while in diffusion equation it appear due to inverted harmonic oscillator   \cite{BhKhLa1995,NoVo1997,BeVi2003,Io2013}
This leads to dilation - contraction operator
in the radial diffusion equation in Ref. \cite{MiRa2018},
where it appears due to a sawtooth field.} \cite{Ross2019}.

\begin{figure}[ht]
\includegraphics[width=0.7\hsize]{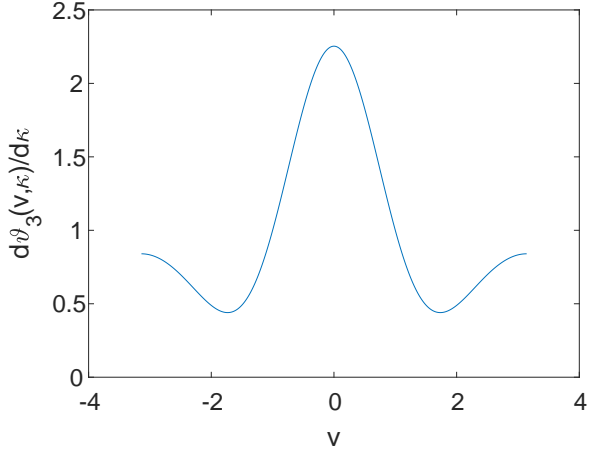}
\caption{An example of derivation of the theta function
with respect to $\kappa\pi$ for $\pi\kappa=0.5$:
$\frac{d\,\vartheta_3(v,\kappa)}{d(\pi\kappa)}$, where
$v=y/R=\phi$.
Here we also use the definition of the theta function
$\vartheta_3(v,\kappa)$ \cite{BaEr55}. }
\label{fig:fig_theta}
\end{figure}

We solve Eq. \eqref{rbc-1} by standard procedures as follows.
Performing the Laplace transformation $\clL[P(t)](s)=\tilde{P}(s)$,
and substituting it in Eq. \eqref{rbc-1}, one has
\begin{subequations}\label{rbc-2}
\begin{align}
\tilde{P}(x,y,s)=e^{-|x|\sqrt{s/D_2}}f(y,s)\label{rbc-2a}\, , \\
D_1\partial_y^2f(y,s)-2\sqrt{sD_2}f(y,s)+\delta(y/R)=0 \label{rbc-2b}\, .
\end{align}
\end{subequations}
Due to the reflecting boundary conditions $f(y=\pm\pi R,s)=0$,
the solution $f(y,s)$ of Eq. \eqref{rbc-2b} is considered as
the superposition
\begin{equation}\label{rbc-3}
f(y,s)=\frac{1}{\sqrt{2\pi R}}\sum_{k=0}^{\infty}f_k(s)\cos(ky/R)\, .
\end{equation}
The initial time backbone dynamics is estimated in \ref{sec:app-A},
and the backbone PDF due to Eq. \eqref{A5} consists of two terms
\begin{multline}\label{rbc-3-4}
f(y,t)=P(x=0,y,t)=\frac{t^{-1/2}}{4\pi R\sqrt{\pi D_2}}
\sum_{n=0}^{\infty}\cos(ny/R) - \\
-\frac{t^{-1/2}D_{\frac{1}{2}}}{4\pi R\sqrt{\pi D_2}}
\sum_{n=1}^{\infty}\cos(ny/R)n^2e^{-\kappa\pi n^2} = \\
=\frac{t^{-1/2}}{4\pi R\sqrt{\pi D_2}}(\delta(y/R)+1)
+\frac{t^{-1/2}D_{\frac{1}{2}}}{4\pi^2 R\sqrt{\pi D_2}}
\frac{d}{d\kappa}\vartheta_3(y/R,\kappa)\, ,
\end{multline}
where $\kappa=2D_{\frac{1}{2}}t^{\frac{1}{2}}/\pi^{3/2}$.
The first term in Eq. \eqref{rbc-3-4} relates to the pining
initial condition, while the second term yields the stationary
solution in the form of the theta function
$\vartheta_3(y/R,\kappa)$ \cite{BaEr55}.  A typical behavior of $\frac{d}{d\kappa}\vartheta_3(y/R,\kappa)$
for $\pi\kappa=0.5$ is shown in Fig. \ref{fig:fig_theta}.

\begin{figure}[ht]
\includegraphics[width=0.9\hsize]{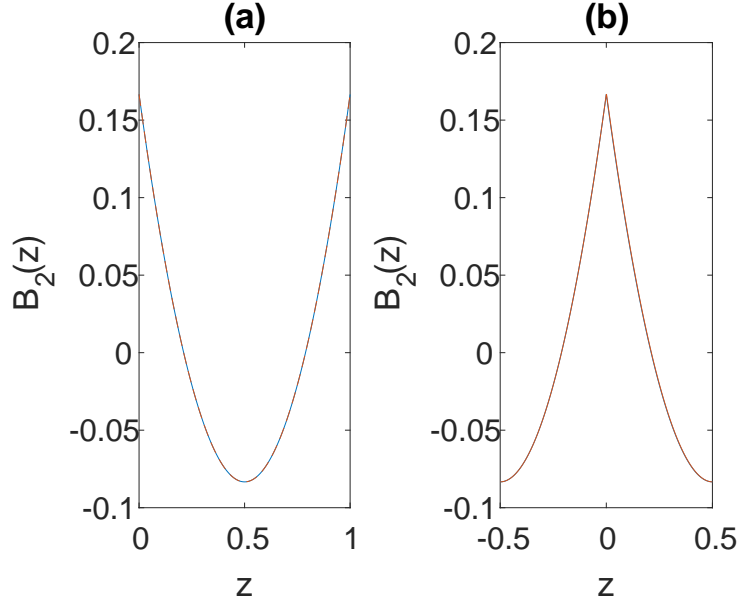}
\caption{Bernoulli polynomial \cite{AbSt72}
$B_2(z)=z^2-z+1/6$ for $z\in [0, \,1]$ on panel (a),
which is also the result of summation in Eq. \eqref{B4}.
Another result of there is of the summation in Eq. \eqref{B4}
for $z\in [-1/2 ,\, 1/2] $ is on panel (b).
It consists of two parts for $z\in [0 ,\, 1/2]$, which coincides with
the left curve of $B_2(z)$. The second part for $z\in [-1/2 ,\, 0]$
corresponds to the curve of the left part of $B_2(z)$ for $z\in [1/2 ,\, 1]$
shifted on $1$.  }
\label{fig:fig_Bernoul1}
\end{figure}

It should admitted that at the large time asymptotic, diffusion in fingers
affects strongly anomalous diffusion in the backbone, and the former should
be taken into account.
Therefore, according to the Laplace inversion, the solution
of Eq. \eqref{rbc-1} reads
\begin{multline}\label{rbc-4}
P(x,y,t)=\frac{1}{\sqrt{2\pi R}}\sum_{k=0}^{\infty}\cos(ky/R)
\clL^{-1}\left[f_k(s)e^{-|x|\sqrt{s/D_2}}\right] = \\
=\frac{1}{\sqrt{2\pi R}}\sum_{k=0}^{\infty}\cos(ky/R)\clL^{-1}
\left[
\frac{e^{-|x|\sqrt{s/D_2}}/\sqrt{D_2}}{2\left(\sqrt{s}
+k^2D_{\frac{1}{2}}\right)} \right] = \\
=\frac{1}{2\pi \sqrt{2RD_2 t}}e^{-\frac{x^2}{4D_2t}}+
\bar{P}(x,y,t)\, ,
\end{multline}
where $D_{\frac{1}{2}}=\frac{D_1}{2\sqrt{D_2}}$. The term $\bar{P}(x,y,t)$
is estimated for the large time in \ref{sec:app-B} and reads
\begin{equation}\label{rbc-5}
\bar{P}(x,y,t)=
\frac{(2-\sqrt{\pi})}{4\pi\sqrt{2 R D_2 t}}
e^{-\frac{x^2}{4D_2t}}\delta(y/R)
+\frac{\pi^{3/2}|x|}{4D_1\sqrt{2 D_2R t^3}}
B_2\left(\frac{y}{2\pi R}\right)e^{-\frac{x^2}{4D_2t}}\, ,
\end{equation}
where $B_2(z)$ is a shifted Bernoulli polynomial \cite{AbSt72}
defined on $z\in (-1/2,\,1/2)$, see Fig. \ref{fig:fig_Bernoul1}.

In the polar coordinates the obtained result in Eqs.
\eqref{rbc-4} and \eqref{rbc-5} reads
\begin{multline}\label{rbc-6}
P(r,\phi,t)=\frac{1}{2\pi \sqrt{2R D_2t}}e^{-\frac{\ln^2(r/R)}{4D_2t}}
\left[1+\frac{2-\sqrt{\pi}}{2}\delta(\phi)\right] + \\
+\frac{\pi^{3/2}|\ln(r/R)|}{4D_1\sqrt{2 D_2R t^3}}
B_2\left(\frac{\phi}{2\pi }\right)e^{-\frac{\ln^2(r/R)}{4D_2t}}\, .
\end{multline}
As obtained, the circular motion is not random due reflections
on boundaries.
It consists of two stationary distributions:
the first one is the initial condition, which relaxes by power law
as $t^{-1/2}$ and the second
stationary distribution is according to
the Bernoulli polynomial $B_2(\phi/2\pi)$, which interacts with the
radial motion. The radial motion is random
and corresponds to the geometric Brownian motion,
which is described by the log-normal distribution \cite{Ross2019},
and leads to the exponential spreading along the radii with the mean
squared displacement (MSD) $\langle r^{2}(t)\rangle\sim e^{t}$.
This dominant process is also
corrected by the L\'evy-Smirnov distribution with respect to
$\frac{t}{\ln(r/R)}$, see \textit{e.g.}, \cite{UcSi13,MeFeHo10}.
Concluding this section it is worth to stress that the geometric Brownian
motion is the geometry effect of the conformal mapping of the
($x-y$) comb model \eqref{rbc-1} onto the circular geometry comb by conformal gluing of the backbone ends.

\section{Periodic boundary conditions}\label{sec:pbc}

\begin{figure}[htbp]
\includegraphics[width=0.4\hsize]{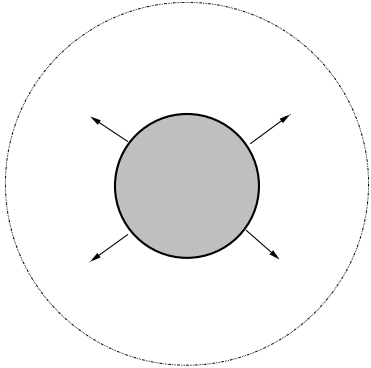}
\caption{Schematic picture of the circular/umbrella comb with the
periodic boundary conditions at $\phi=\pm\pi$ and  outward
diffusion along the radii with zero boundary conditions at infinity.}
\label{fig:fig_pbc}
\end{figure}
In the section, we consider the second scenario with periodic circular
motion at $r=R$ and the \textit{outward} radial motion with a
constant diffusion coefficient $D_r=D$, see Fig.~\ref{fig:fig_pbc}.
Equation \eqref{int-uc-1} now is
\begin{equation}\label{pbc-1}
\partial_tP=D_1\delta(r-R)\frac{1}{r^2}\partial_{\phi}^2P
+D\frac{1}{r}\partial_r r\partial_rP\, .
\end{equation}
The initial condition is $P_0=P(r,\phi,t=0)=\delta(r-R)\delta(\phi)$
The distribution function $P=P(r,\phi,t)$  is a convolution integral
\begin{equation} \label{pbc-2}
P(r,\phi,t)=\int_0^tG(r,t-t')F(\phi,t')dt'\, ,
\end{equation}
which represents two independent motions in the Laplace space:
\begin{equation}\label{pbc-3}
\tilde{P}(r,\phi,s)=g(r,s)f(\phi,s)\, .
\end{equation}
This corresponds to Eq. \eqref{pbc-1} in the Laplace space,
\begin{equation}\label{pbc-4}
s\tilde{P}-P_0=D_1\delta(r-R)\frac{1}{r^2}\partial_{\phi}^2\tilde{P}
+D\frac{1}{r}\partial_r r\partial_r\tilde{P}\, .
\end{equation}
Now, the boundary conditions for the radial motion can be easily
specified for $\tilde{P}$, $g$ and $f$. These are
$ \tilde{P}(r=R,\phi,s)=f(\phi,s)$ and $g(r=R,s)=1$, and
$\tilde{P}(r=\infty,\phi,s)=g(r=\infty,s)=
\tilde{P}'(r=\infty,\phi,s)=g'(r=\infty,s)=0$, where prime
means differentiation with respect to $r$.
Note also, $f(\phi,s)$ is a multivalued function, and
the conformal map cannot be performed. Therefore, we are treating
the problem in the polar coordinates.

First, let us consider diffusion in radii - fingers. In the Laplace space,
the diffusion equation from Eq. \eqref{pbc-4} leads to the equation
\begin{equation}\label{pbc-5}
rsg=Drg''+Dg'
\end{equation}
with the solution
\begin{equation}\label{pbc-6}
g(r,s)=A(s)K_0\left(r\sqrt{s/D}\right)\theta(r-R)\, ,
\end{equation}
where $K_0(z)$ is the modified Bessel function of the second kind,
which satisfies the boundary conditions at infinity. The second boundary
condition $g(R,s)=1$ yields
$A(s)=2\left[K_0\left(R\sqrt{s/D}\right)\right]^{-1}$,
where $\theta(0)=1/2$ is accounted.

Substituting solution \eqref{pbc-6} in Eq. \eqref{pbc-4},
one obtains, see \ref{sec:app-C}
\begin{equation}\label{pbc-7}
\partial_{\phi}^2f+af-bs^{\frac{1}{2}}f+c\delta(\phi)=0\, ,
\end{equation}
where $a=2RD/D_1$ and $bs^{1/2}=
\frac{2R^2K_1(\lambda R)}{D_1K_0(\lambda R)} \sqrt{Ds}$,
and $c=R^2/D_1$.

\subsection{Initial time asymptotics}

For the initial times, when $s\rightarrow \infty$, and
$\lambda R=R\sqrt{s/D}\gg 1$, one obtains
$\frac{K_1(\lambda R)}{K_0(\lambda R)}\approx 1$ and
$a\ll bs^{\frac{1}{2}}$. Then, Eq. \eqref{pbc-7} is simplified
with the solution
\begin{equation}\label{pbc-8}
f(\phi,s)=\sum_{n=-\infty}^{\infty}\frac{ce^{i n\phi}}{n^2+b\sqrt{s}}\, .
\end{equation}
We also obtain that $g(r,s)\approx R^{\half}e^{-\lambda(r-R)}/\sqrt{r}$ \cite{AbSt72} and the PDF $P(r,\phi,t)$ leads to a chain of estimations
in \ref{sec:app-C}. Therefore, the PDF reads
\begin{subequations}\label{pbc-9}
\begin{align}
P(r,\phi,t)=  & \clL^{-1}\left[g(r,s)f(\phi,s)\right] \approx
\sum_{n=-\infty}^{\infty}\frac{cR^{\half}}{br^{\half}}e^{i n\phi}
\cdot
\clL^{-1}\left[\frac{e^{-\frac{\sqrt{s}(r-R)}{\sqrt{D}}}}
{n^2/b+\sqrt{s}}\right] = \label{pbc-9a} \\
= &
\frac{cR^{\half}}{b\sqrt{r\pi t}}e^{-\frac{(r-R)^2}{4Dt}}[1+2\delta(\phi)]
-\frac{2c(Dt/r)^{\half}}{b(r-R)}e^{-\frac{(r-R)^2}{4Dt}}  + \\
+ & \frac{c\pi\sqrt{RD^{\half}}}{\sqrt{rb(r-R)}}
\frac{\cosh\left[(\pi-|\phi|)\sqrt{D^{\half}t/b(r-R)}\right]}
{\cosh\left[\pi\sqrt{D^{\half}t/b(r-R)}\right]} e^{-\frac{(r-R)^2}{4Dt}}
\label{pbc-9b} \, ,
\end{align}
\end{subequations}
where $\phi\in[-\pi\, , \pi]$. The modulus $|phi|$ is due to the symmetry
of Eq. \eqref{pbc-7}.
We also stress that the solution  \eqref{pbc-9b}
is valid for $r>R$, strictly.
It should be noted that for $t\rightarrow 0$, the solution
reduces to the transport for $r=R$, and  Eq. \eqref{pbc-9a} reads
\begin{multline}\label{pbc-9-2}
P(R,\phi,t)
\approx
\frac{c}{b\sqrt{\pi t}}[1+2\delta(\phi)]
-\frac{2c}{\sqrt{t}}
\sum_{n=1}^{\infty}\frac{\cos(n\phi)}{n^2} = \\
\frac{c}{b\sqrt{\pi t}}[1+2\delta(\phi)]
-\frac{2c\pi^2}{\sqrt{t}}B_2\left(\frac{\phi}{2\pi}\right) \, .
\end{multline}
The situation changes dramatically for the long times.

\subsection{Large time asymptotics}

For the large times, when $\lambda R\ll 1$, we have
 $K_0(\lambda R)\approx \ln\frac{2}{\gamma\lambda R}$ and
$K_1(\lambda R) \approx 1/\lambda R$,
where $\gamma$ is the Euler constant \cite{AbSt72}.
However,  asymptotic behavior of $K_0(\lambda r)$
must correspond to the boundary conditions at $r\rightarrow\infty$.
Therefore, we take the intermediate asymptotic, when $\lambda r>1$,
which yields
$$K_0(\lambda R)\approx \sqrt{\frac{\pi}{2\lambda r}}e^{-\lambda r}=
\sqrt{\pi (D/s)^{\half}/r }e^{-r(s/D)^{\half}}\, .$$
Correspondingly, in the limits $s\rightarrow 0$ and
$rs^{\half}D^{-\half} >1$, the radial distribution
$g(r,s)$ and the coefficient $b$ in Eq. \eqref{pbc-7} are functions of $s$
which are approximated as follows
\begin{subequations}\label{pbc-10}
\begin{align}
& g(r,s)\approx \sqrt{2\pi D^{\half}/rs^{\half}}
\frac{e^{-r(s/D)^{\half}}}
{\ln\left(4D/\gamma^2 R^2s\right)}\, , \label{pbc-11a}  \\
& bs^{\half}=b(s)s^{\half}\approx \frac{4DR}{D_1}
\ln^{-1}\left(4D/\gamma^2 R^2 s\right)
\equiv b_1\left[\ln\left(4D/\gamma^2R^2s\right)\right]^{-1}
\label{pbc-11b} \, .
\end{align}
\end{subequations}

\begin{figure}[ht]
\includegraphics[width=0.7\hsize]{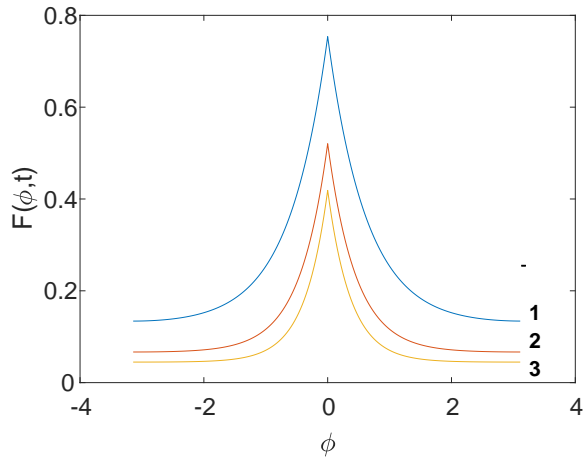}
\caption{Evolution of the braces $F(\phi,t)$ in PDF of Eq. \eqref{pbc-13}.
The plots are dependence of $F(\phi,t)$ on $\phi$ for three
times where plot 1 corresponds to $t=10$, 2 to $t=100$ and 3 to $t=1000$.
All parameters are taken to be one, while $\gamma=1.781$ }
\label{fig:fig_Backbone}
\end{figure}

Again neglecting the parameter $a$ in Eq. \eqref{pbc-7}, we obtain
the solution as follows
\begin{equation}\label{pbs-11}
f(\phi,s)=\sum_{n=-\infty}^{\infty}
\frac{ce^{i n\phi}\ln\left(D/\gamma^2R^2s\right)}
{b_1 +n^2\ln\left(D/\gamma^2R^2s\right)}\, .
\end{equation}
Then the PDF \eqref{pbc-9} for the large time asymptotics reads
in the form of the inverse Laplace transformation
\begin{multline}\label{pbc-12}
P(r,\phi,t)= \clL^{-1}\left[g(r,s)f(\phi,s)\right] \approx \\
\approx \clL^{-1}\left[\sum_{n=-\infty}^{\infty}
\frac{2ce^{i n\phi}}{n^2+b_1\ln\left(4D/\gamma^2R^2s\right)}
\cdot
\sqrt{2\pi (D/r^2s)^{\half}} e^{-r(s/D)^{\half}}
\right] \, .
\end{multline}
The solution, obtained in \ref{sec:app-C}, reads
\begin{multline}\label{pbc-13}
P(r,\phi,t)\approx  \frac{c\pi}{b_1\Gamma\left(\frac{1}{4}\right)}
\frac{\sqrt{2\pi (D/t^3r^2)^{\half}}}
{\ln^{\half}\left(4Dt/\gamma^2R^2\right)} e^{-3r^2/8Dt}\\
\times\Biggl\{\frac{\ln^{-\half}\left(4Dt/\gamma^2R^2\right)}{\pi}+
\frac{\cosh\left[(\pi-|\phi|)\sqrt{b_1\ln\left(4Dt/\gamma^2R^2\right)}\right]}
{\cosh\left[\pi\sqrt{b_1\ln\left(4Dt/\gamma^2R^2\right)}\right]}\Biggr\}
\, .
\end{multline}
This result is valid for $r\gg R$ and $t\rightarrow\infty$,
and the Tauberian theorem, applied in Eq. \eqref{C10},
grasps exactly this intermediate asymptotic behavior of
$K_0(\lambda r)\propto r^{-\half}s^{-\frac{1}{4}}$ due to the power law.
The logarithmic evolution in the backbone shown
in Fig.~\ref{fig:fig_Backbone}, relaxes to the radii-fingers.
However, in large time asymptotic calculation this backbone relaxation
cannot be separated from the radii one. Note also that the approximate solution \eqref{pbc-13} for the PDF in the specific area does not conserve the probability $\clP$, namely the latter is
\begin{equation}\label{pbc-14}
\clP(t)=\int_0^{\infty}dr\int_{-\pi}^{\pi}d\phi
P(r,\phi,t)\sim t^{-\frac{1}{2}}\, .
\end{equation}

\subsection{The MSD in radii}

This also results to radii subdiffusion with
the mean squared displacement (MSD) of the order
of $\langle r^2(t)\rangle \sim Dt$.
Indeed, the relaxation process in the backbone contributes
to the transport in fingers with the MSD defined from the inverse Laplace transformation in Eq. \eqref{pbc-12} as follows
\begin{multline}\label{C13}
\langle r^2(t)\rangle =\clP^{-1}(t)\clP^{-1}(t)\frac{1}{2\pi}\int_R^{\infty}\int_{-\pi}^{\pi}
r^2P(r,\phi,t)dr d\phi= \\
=\clL^{-1}\left[
\frac{2c}{b_1\ln\left(4D/\gamma^2R^2s\right)}
\cdot\sqrt{2\pi (D/s)^{\half}}
\int_R^{\infty} r^{3/2} e^{-r(s/D)^{\half}}dr\right]  \propto \\
\propto D\clP^{-1}(t)\clL^{-1}\Big[ s^{-3/2}
\Gamma\left(\frac{3}{2},R\sqrt{s}\right)\Big](t) \, ,
\end{multline}
where $\Gamma(\alpha,z)$ is the incomplete gamma function \cite{AbSt72}
and the factor $\clP^{-1}(t)$ is due to the non conserved probability.
Performing the Laplace inversion by means of the Tauberian theorem,
see \ref{sec:app-C},
we obtain a subdiffusive growth of the order of $Dt^{1/2}$
However, the obtained expression
should be normalized by the probability $\clP(t)$.
This eventually yields the MSD in the form of normal diffusion
\[ \langle r^2(t)\rangle \sim D t\, .\]

\section{Conclusion}

A circular comb is considered, and
two scenarios of the anomalous transport in the
circular comb geometry are studied. The first scenario
corresponds to the conformal mapping of a
comb Fokker-Planck equation on the umbrella comb.
In this case, the reflecting boundary conditions are
imposed on the circular (rotator) motion, while the radial
motion corresponds to geometric Brownian motion
with vanishing to zero boundary conditions on infinity.
The radial diffusion is described by
the log-normal distribution, which corresponds to exponentially fast
motion with the MSD of the order of $e^t$.
The second scenario corresponds to circular diffusion
with periodic boundary conditions and the outward Brownian radial
diffusion with vanishing to zero boundary conditions at infinity.
In this case the radial motion is normal diffusion with
the MSD of the order of $t$.
However the circular motion in both scenarios is a superposition
of cosine functions that results in a stationary distribution in the
form of the Bernoulli polynomials, with the power law relaxation.

\appendix

\section{Backbone dynamics with RBC}\label{sec:app-A}

\def\theequation{A.\arabic{equation}}
\setcounter{equation}{0}

Note that ``RBC'' means reflecting boundary conditions
at $y=\pm \pi R$.

Let us consider Eq. \eqref{rbc-2b}
\begin{equation}\label{A1}
D_1\partial_y^2f(y,s)-2\sqrt{sD_2}f(y,s)+\delta(y/R)=0\, ,
\end{equation}
which describes anomalous diffusion, namely subdiffusion,
along the  $y$-axis-backbone. Recall that it also corresponds to
anomalous diffusion in the ring backbone of the umbrella
comb.
Due to the reflection boundary conditions $f'(y=\pm \pi R,s)=0$,
the solution can be presented as the superposition of the
even eigenfunctions $\frac{1}{\sqrt{2\pi R}}\cos(ny/R)$. It reads
\begin{equation}\label{A2}
f(y,s)=\frac{1}{\sqrt{2\pi R}}\sum_{n=0}^{\infty}f_n(s)\cos(ny/R)\, .
\end{equation}
From Eq. \eqref{A1}, this yields
\begin{equation}\label{A3}
f_n(s)=\frac{1}{2\sqrt{2\pi R D_2}}\frac{1}{\sqrt{s}+D_{\frac{1}{2}}n^2}\, ,
\end{equation}
where $D_{\frac{1}{2}}=D_1/2\sqrt{D_2}$ is a subdiffusion coefficient.
Performing the inverse Laplace transform, one obtains
the solution in the form of the Mittag-Leffler function \cite{BaEr55}
\begin{multline}\label{A4}
f_n(t)=\frac{1}{2\sqrt{2\pi R D_2}}\frac{1}{2\pi i}\int_{-i\infty}^{i\infty}
\frac{e^{st}}{\sqrt{s}+D_{\frac{1}{2}}n^2}ds= \\
=\frac{t^{-1/2}}{2\sqrt{2\pi R D_2}}E_{\frac{1}{2},\frac{1}{2}}
\left(-n^2D_{\frac{1}{2}}t^{\frac{1}{2}}\right)\, .
\end{multline}
Using properties of the Mittag-Leffler functions \cite{BaEr55}
\begin{align*}
E_{\alpha,1}(z)=E_{\alpha}(z) \quad \text{and}
\quad E_{\alpha,\beta}(z)=\frac{1}{\Gamma(\beta)}+
zE_{\alpha, \alpha+\beta}(z)\, , \\
E_{\alpha,\beta}(z)=\sum_{k=0}^{\infty}\frac{z^k}{\Gamma(k\alpha+\beta)}\, ,
\end{align*}
where $\Gamma(\nu+1)=\nu\Gamma(\nu)$ is a gamma function,
we obtain the initial behavior
of $f_n(t)$ in Eq. \eqref{A4}:
\begin{multline}\label{A5}
f_n(t)=\frac{1}{2\sqrt{2\pi R D_2}}\left[\frac{t^{-1/2}}{\Gamma(1/2)}
-n^2D_{\frac{1}{2}}E_{\frac{1}{2}}\left(-n^2D_{\frac{1}{2}}t^{\frac{1}{2}}\right)
\right] = \\
=\frac{1}{2\sqrt{2\pi R D_2}}\left\{\frac{t^{-1/2}}{\Gamma(1/2)}
-n^2D_{\frac{1}{2}} \cdot
\exp\left[-n^2D_{\frac{1}{2}}t^{\frac{1}{2}}/\Gamma(3/2)\right]\right\}\, ,
\end{multline}
where $\Gamma(1/2)=\sqrt{\pi}$.
Taking into account the expansion \eqref{A2}, we obtain
the first term in Eq. \eqref{A5} in the form of the pining initial condition
decaying with time, while the second term is the theta function
$\vartheta_3(y/R,\kappa)$.  Then Eq. \eqref{A2} reads
\begin{multline}\label{A6}
f(y,t)=P(x=0,y,t)=\frac{t^{-1/2}}{4\pi R\sqrt{\pi D_2}}
\sum_{n=0}^{\infty}\cos(ny/R) - \\
-\frac{t^{-1/2}D_{\frac{1}{2}}}{4\pi R\sqrt{\pi D_2}}
\sum_{n=1}^{\infty}\cos(ny/R)n^2e^{-\kappa\pi n^2} = \\
=\frac{t^{-1/2}}{4\pi R\sqrt{\pi D_2}}(\delta(y/R)+1)
+\frac{t^{-1/2}D_{\frac{1}{2}}}{4\pi^2 R\sqrt{\pi D_2}}
\frac{d}{d\kappa}\vartheta_3(y/R,\kappa)\, ,
\end{multline}
where $\kappa=
2D_{\frac{1}{2}}t^{\frac{1}{2}}/\pi^{3/2}$.
Here we also use the definition of the theta function
$\vartheta_3(y/R,\kappa)$ \cite{BaEr55}.
A typical behavior of $\frac{d}{d\kappa}\vartheta_3(y/R,\kappa)$
for $\pi\kappa=0.5$ is shown in Fig. \ref{fig:fig_theta}.

\section{Long time asymptotic}\label{sec:app-B}

\def\theequation{B.\arabic{equation}}
\setcounter{equation}{0}

The long time diffusion can be estimated
from Eq. \eqref{A4}, as well.  To that end we take into account that
the Laplace inversion of the Mittag-Leffler function can be
presented in the form of the error function $\mathrm{Erfc}(z)$
\cite{BaEr54}, as follows
\begin{equation}\label{Error-function}
E_{\frac{1}{2},\frac{1}{2}}(at^{\frac{1}{2}})=
\frac{1}{\sqrt{\pi t}}-
a e^{a^2t}\mathrm{Erfc}\left(at^{\frac{1}{2}}\right)\, .
\end{equation}
However, at the large time asymptotics, diffusion in fingers
affects strongly anomalous diffusion in the backbone,
and the former should
be taken into account. Therefore, we consider
the inverse Laplace transformation in Eq. \eqref{rbc-4}, which
is the table integral \cite{BaEr54} of the form
\begin{multline}\label{B1}  
\sum_{n=0}^{\infty}\cos(ny/R)\clL^{-1}\left[\frac{e^{-a\sqrt{s}}}{\sqrt{s}+b}\right]
=(\pi t)^{-1/2}e^{-a^2(4t)^{-1}}-\sum_{n=1}^{\infty}\cos(ny/R) \\
\times\left[(\pi t)^{-1/2}e^{-a^2(4t)^{-1}}-
be^{ba+b^2t}\mathrm{Erfc}\left(2^{-1}at^{-1/2}+bt^{1/2}\right)\right]\, .
\end{multline}
Here parameters $a$ and $b$ are determined from Eqs. \eqref{rbc-4}
and \eqref{A4}. Namely, $a=|x|/\sqrt{D_2}$ determines radial diffusion,
while $b=D_{\frac{1}{2}}n^2$ specifies backbone subdiffusion as
obtained above. The error function in Eq. \eqref{B1}
for the large argument reads
\begin{multline}\label{B2}  
\mathrm{Erfc}\left(2^{-1}at^{-1/2}+bt^{1/2}\right)\approx
\frac{e^{-\frac{a^2}{4t}-ab -b^2t}}{at^{-1/2}+2bt^{1/2}} \approx \\
\approx
e^{-\frac{a^2}{4t}-ab -b^2t}
\left[(2b)^{-1}t^{-1/2}-a(2b)^{-2}t^{-3/2}\right] \, .
\end{multline}
Taking into account expansion \eqref{A2} and Eqs. \eqref{rbc-4},
\eqref{B1}, and \eqref{B2}, and values of $a$ and $b$, we obtain
\begin{multline}\label{B3}  
P(x,y,t)=
\frac{1}{2\sqrt{2\pi D_2 R}}\sum_{n=0}^{\infty}\cos(ny/R)
\clL^{-1}\left[\frac{e^{-a\sqrt{s}}}{\sqrt{s}+b}\right] = \\
=\frac{1}{2\pi\sqrt{2 D_2 t}}e^{\frac{x^2}{4D_2t}}+\bar{P}(x,y,t) = \\
=\frac{1}{2\pi\sqrt{2 R D_2 t}}
e^{-\frac{x^2}{4D_2t}}\left[1+\delta(y/R)\right]
-\frac{1}{4\sqrt{2\pi R D_2 t}}e^{-\frac{x^2}{4D_2t}}\delta(y/R) \\
+\frac{|x|}{4D_1\sqrt{2\pi D_2R t^3}}
e^{-\frac{x^2}{4D_2t}}
\sum_{n=1}^{\infty} \frac{\cos(ny/R)}{n^2} \, .
\end{multline}
Estimating the last term in Eq. \eqref{B3}, which is \cite{PrBrMa86}
\begin{equation}\label{B4}   
\sum_{n=1}^{\infty}\frac{\cos(ny/R)}{n^2}=
\pi^2B_2\left(\frac{y}{2\pi R}\right)\, ,
\end{equation}
we obtain
\begin{equation}\label{B5}  
\bar{P}(x,y,t)=
\frac{(2-\sqrt{\pi})}{4\pi\sqrt{2 R D_2 t}}
e^{-\frac{x^2}{4D_2t}}\delta(y/R)
+\frac{\pi^{3/2}|x|}{4D_1\sqrt{2 D_2R t^3}}
B_2\left(\frac{y}{2\pi R}\right)e^{-\frac{x^2}{4D_2t}}\, .
\end{equation}
Here $B_2(z)$ is a shifted Bernoulli polynomial presented
on $z\in (-1/2,\,1/2)$, see Fig. \ref{fig:fig_Bernoul1}.
It should be admitted that for the standard
definition of the Bernoulli polynomial,
$B_2(z)=z^2-2z+1/6$ is defined on $z\in (0,1)$ \cite{AbSt72}

\section{Backbone dynamic with PBC}\label{sec:app-C}
\def\theequation{C.\arabic{equation}}
\setcounter{equation}{0}

Note that ``PBC'' means periodic boundary conditions
at $\phi=\pm \pi$.

Let us consider the radial motion in Eq. \eqref{pbc-4}
according to the radial derivative
$\frac{1}{r}\partial_r r\partial_rg=g^{''}+r^{-1}g'$, where
$g=g(r,s)$ is the solution of Eq. \eqref{pbc-5}
\begin{equation}\label{C1}
g(r,s)=\frac{2K_0\left(\lambda r\right)}{K_0\left(\lambda R\right)}
\theta(r-R)\, ,\quad \lambda=\sqrt{s/D}\,
\end{equation}
in the form is the modified Bessel function of the second kind
$K_0(z)$, which satisfies the boundary conditions at infinity.
Using properties of its derivatives \cite{AbSt72} as follows
\begin{subequations}\label{C2}
\begin{align}
K_0^{\prime}(z) & =-K_1(z) \, , \label{C2a} \\
K_0^{\prime\prime}(z) & =-K_1^{\prime}(z)=
\frac{1}{2}\left[K_0(z)+K_2(z)\right]\, , \label{C2b} \\
-K_1^{\prime}(z) & =K_0(z)+\frac{1}{z}K_1(z) \, , \label{C2c}
\end{align}
\end{subequations}
we have
\begin{equation}\label{C3}
g'(r,s)=\frac{2}{K_0(\lambda R)}\left[
-\lambda K_1(\lambda r)\theta(r-R)+K_0(\lambda r)\delta(r-R)\right]
\end{equation}
and
\begin{multline}\label{C4}
g''(r,s)=\frac{2}{K_0(\lambda R)}\left[\lambda^2K_0(\lambda r)
\theta(r-R) + \right.\\ \left.
+\lambda r^{-1}K_1(\lambda r)\theta(r-R)-\lambda K_1(\lambda r)\delta(r-R)
\right] \, ,
\end{multline}
where $-K_0(\lambda r)\delta'(r-R)=\lambda K_1(\lambda r)\delta(r-R)$ is
used. Therefore, the superposition of Eqs. \eqref{C3}
and \eqref{C4} yields
\begin{multline}\label{C5}
g''(r,s)+\frac{1}{r} g'(r,s)=
\frac{2}{K_0(\lambda R)} \left[\lambda^2K_0(\lambda r)\theta(r-R)+\right. \\
\left.+\frac{1}{R}K_0(\lambda r)\delta(r-R) -
\lambda K_1(\lambda r)\delta(r-R)\right]\, .
\end{multline}

Inserting the obtained result \eqref{C5} in Eq. \eqref{pbc-4}
and taking into account  Eq. \eqref{pbc-3}, we
obtain the equation for $f(\phi,s)$  as follows
\begin{equation}\label{C6}
D_1\frac{1}{R^2}\partial_{\phi}^2f
+\left[\frac{2D}{R}-\frac{2K_1(\lambda R)}{K_0(\lambda R)} \sqrt{Ds}
\right] f+\delta(\phi)=0 \, .
\end{equation}

\subsection{Initial time asymptotics}

For the initial times, when $s\rightarrow \infty$, and
$\lambda R=R\sqrt{s/D}\gg 1$, it is obtained that
$f(\phi,s)$ is determined by Eq. \eqref{pbc-8}.
Taking into account the solution \eqref{pbc-8} and accounting
that $g(r,s)\approx R^{\half}e^{-\lambda(r-R)}/\sqrt{r}$ \cite{AbSt72},
we obtain for the PDF $P(r,\phi,t)$ the following chain of estimations
(see also \ref{sec:app-B})
\begin{equation}\label{C7a}
P(r,\phi,t)= \clL^{-1}\left[g(r,s)f(\phi,s)\right] \approx
\clL^{-1}\left[
\sum_{n=-\infty}^{\infty}\frac{cR^{\half}}{br^{\half}}e^{i n\phi}
\cdot\frac{e^{-\frac{\sqrt{s}(r-R)}{\sqrt{D}}}}{n^2/b+\sqrt{s}}\right]\, .
\end{equation}
Performing the Laplace inversion, separating term with $n=0$,
and accounting Eqs. \eqref{B1}, \eqref{B2}, we obtain
\begin{multline}\label{C7b}
P(r,\phi,t) = \frac{cR^{\half}}{br^{\half}\sqrt{\pi t}}
e^{-\frac{(r-R)^2}{4Dt}}[1+2\delta(\phi)]
-\sum_{n=1}^{\infty}\frac{2cR}{br}\cos(n\phi)
 \\
\times e^{n^2(r-R)b^{-1}D^{-\half}+b^{-2}n^4t}
\mathrm{Erfc}\left(2^{-1}(r-R)D^{-\half}t^{-1/2}+n^2b^{-1}t^{1/2}\right)
\approx \\
\approx
\frac{cR^{\half}}{b\sqrt{\pi r t}}e^{-\frac{(r-R)^2}{4Dt}}[1+2\delta(\phi)]
-\frac{2cR^{\half}}{\sqrt{rt}}
\sum_{n=1}^{\infty}\frac{\cos(n\phi)e^{-\frac{(r-R)^2}{4Dt}}}
{2^{-1}b(r-R)D^{-\half}t^{-1}+n^2} = \\
=\frac{cR^{\half}}{b\sqrt{r\pi t}}e^{-\frac{(r-R)^2}{4Dt}}[1+2\delta(\phi)]
-\frac{2c(RDt/r)^{\half}}{b(r-R)}e^{-\frac{(r-R)^2}{4Dt}}  + \\
+\frac{c\pi\sqrt{RD^{\half}}}{\sqrt{rb(r-R)}}
\frac{\cosh\left[(\pi-|\phi|)\sqrt{D^{\half}t/b(r-R)}\right]}
{\cosh\left[\pi\sqrt{D^{\half}t/b(r-R)}\right]} e^{-\frac{(r-R)^2}{4Dt}}\, ,
\end{multline}
where $\phi\in[-\pi\, , \pi]$. Note, that the modulus
$|phi|$ is due to the symmetry of Eq. \eqref{pbc-7}.
We also stress that the solution is valid for
$r>R$, strictly. For $r=R$, Eq. \eqref{C7b} reads
\begin{multline}\label{C7c}
P(r,\phi,t)
\approx
\frac{c}{b\sqrt{\pi t}}[1+2\delta(\phi)]
-\frac{2c}{\sqrt{t}}
\sum_{n=1}^{\infty}\frac{\cos(n\phi)}{n^2} = \\
=\frac{c}{b\sqrt{\pi t}}[1+2\delta(\phi)]
-\frac{2c\pi^2}{\sqrt{t}}B_2\left(\frac{\phi}{2\pi}\right) \, .
\end{multline}


\subsection{Large time asymptotics }
In this section we estimate Eq. \eqref{pbc-12}.
Performing the inverse Laplace transformation in Eq.
\eqref{pbc-12}, we have
\begin{multline}\label{C8}
P(r,\phi,t)= \clL^{-1}\left[g(r,s)f(\phi,s)\right] \approx \\
\approx \clL^{-1}\left[\sum_{n=-\infty}^{\infty}
\frac{2ce^{i n\phi}}{n^2+b_1\ln\left(4D/\gamma^2R^2s\right)}
\cdot
\sqrt{2\pi (D/r^2s)^{\half}} e^{-r(s/D)^{\half}}
\right] \, .
\end{multline}
Performing the Laplace inversion term by term
in the summation $\sum_n A_n(s)$,
we have
\begin{equation}\label{C9}
\clL^{-1}\left[A_n(s)\right](t)
= \clL^{-1}\left[\frac{d_1s^{-1/4}}{n^2+S(s)}e^{-d_2\sqrt{s}}\right]\, ,
\end{equation}
where $S(s)=b_1\ln\left(4D/\gamma^2R^2s\right)$ and
$e^{-d_2\sqrt{s}}\sim \left(1+d_2s^{\half}\right)^{-1}$.
Making scaling by $\lambda$, we obtain
from Eq. \eqref{C9} that $A_n(\lambda s)/A_n(s)\sim \lambda^{-1/4}$
for $s\rightarrow 0$, while $[n^2+S(s)]^{-1}$ is
slow functions of $1/s$. Therefore,
$A_n(s)=s^{-\rho}e^{-d_2\sqrt{s}} L(1/s)$,
where $L(1/s)=[n^2+S(s)]^{-1} $ is a slow function of $1/s$ and $\rho=1/4$.
Then applying the Tauberian theorem, we obtain \cite{Fe71}
$A_n(t)=t^{\rho-1} e^{-3d_2^2/8t}  L(t)/\Gamma(\rho)$, see Sec.
\ref{sec:tau} below.
Expression \eqref{C8} reads now as follows
\begin{equation}\label{C10}
P(r,\phi,t)\approx \sqrt{2\pi (D/t^3r^2)^{\half}} e^{-3r^2/8Dt}
\sum_{n=-\infty}^{\infty}
\frac{2ce^{i n\phi}}{\Gamma(1/4)[n^2+S(t)]}\, ,
\end{equation}
where $S(t)=b_1\ln\left(4Dt/\gamma^2R^2\right)$.
The summation yields \cite{PrBrMa86}
\begin{multline}\label{C11}
\sum_{n=-\infty}^{\infty}
\frac{2ce^{i n\phi}}{n^2+S(t)}=
\frac{2c}{S(t)}+\sum_{n=1}^{\infty}\frac{2c\cos(n\phi)}{n^2+S(t)} = \\
=\frac{2c}{S(t)}+\frac{c\pi}{\sqrt{S(t)}}\cdot
\frac{\cosh\left[(\pi-|\phi|)\sqrt{S(t)}\right]}
{\cosh\left[\pi\sqrt{S(t)}\right]}-\frac{c}{S(t)} \, .
\end{multline}
Eventually, one obtains the PDF for the large time asymptotics as follows
\begin{multline}\label{C12}
P(r,\phi,t)\approx  \frac{c\pi}{b_1\Gamma\left(\frac{1}{4}\right)}\cdot
\frac{\sqrt{2\pi (D/t^3r^2)^{\half}}}
{\ln^{\half}\left(4Dt/\gamma^2R^2\right)} e^{-3r^2/8Dt}\\
\times\Biggl\{\frac{\ln^{-\half}\left(4Dt/\gamma^2R^2\right)}{\pi}+
\frac{\cosh\left[(\pi-|\phi|)\sqrt{b_1\ln\left(4Dt/\gamma^2R^2\right)}\right]}
{\cosh\left[\pi\sqrt{b_1\ln\left(4Dt/\gamma^2R^2\right)}\right]}\Biggr\}
\, .
\end{multline}

\subsection{The Tauberian theorem}\label{sec:tau}
Let us consider the Laplace inversion of
$A_n(s)=s^{-\rho}e^{-d_2\sqrt{s}}$ with $\rho=1/4$.
To this end let us consider the exponential in the form of the
Fourier transform
\begin{equation}\label{tau1}
\int_{-\infty}^{\infty}\frac{e^{ikd_2}dk}{(k+is^{\half})(k-is^{\half})}=
\frac{\pi e^{-d_2\sqrt{s}}}{\sqrt{s}}\, ,  \quad d_2>0\, .
\end{equation}
Since the PDF is a real function, thus we take the real part of
the integrations to avoid artificial complex parts that
can appear due to this trick.
Therefore, we have the
\begin{multline}\label{tau2}
A_n(t)=\mathrm{Re}\clL^{-1}[A_n(s)](t)=\frac{1}{2\pi^2 i}
\int_{-\infty}^{\infty}e^{ikd_2}
\int_C\frac{s^{\frac{1}{4}}e^{st}}{s+k^2}dsdk = \\
=\mathrm{Re}\frac{t^{-\frac{1}{4}}}{\pi}\int_{-\infty}^{\infty}
e^{-k^2t+ikd_2}k^{\half}dk =
\frac{t^{-\frac{3}{4}}}{\pi}\int_{0}^{\infty}
e^{-z^2+izkd_2}z^{\half}dz = \\
=\frac{(4t)^{-\frac{3}{4}}}{\pi}\Gamma(\frac{3}{2})
\Psi\left(\frac{3}{4},\half;-\frac{d_2^2}{4t}\right)\approx \\
\approx \frac{(4t)^{-\frac{3}{4}}}{\pi}
\frac{\Gamma(\frac{3}{2})\Gamma(\half)}{\Gamma(\frac{5}{4})}
\left(1-\frac{3d_2^2}{8t}\right)\propto t^{-\frac{3}{4}}
e^{-\frac{3d_2^2}{8t}}\, .
\end{multline}
Here $\Psi(a,s;z)$ is the degenerate hypergeometric function,
obtained in the second integration with respect to $z$,
which is a table integral \cite{PrBrMa86}.

\end{document}